\documentclass{article}

\usepackage{arxiv}

\usepackage[utf8]{inputenc} 
\usepackage[T1]{fontenc}    
\usepackage{hyperref}       
\usepackage{url}            
\usepackage{booktabs}       
\usepackage{amsfonts}       
\usepackage{nicefrac}       
\usepackage{microtype}      
\usepackage{lipsum}		
\usepackage{graphicx}
\usepackage{natbib}
\usepackage{doi}

\usepackage{cite}
\usepackage{comment}
\usepackage{amsmath,amssymb,amsfonts}
\usepackage{algorithmic}
\usepackage{graphicx}
\usepackage{textcomp}
\usepackage{multirow}
\usepackage{dblfloatfix}
\usepackage{booktabs,siunitx}
\usepackage[symbol]{footmisc}
\usepackage{tablefootnote}
\usepackage{enumitem}

\def\BibTeX{{\rm B\kern-.05em{\sc i\kern-.025em b}\kern-.08em
    T\kern-.1667em\lower.7ex\hbox{E}\kern-.125emX}}
\usepackage{array}
\newcolumntype{L}[1]{>{\raggedright\let\newline\\\arraybackslash\hspace{0pt}}m{#1}}
\newcolumntype{C}[1]{>{\centering\let\newline\\\arraybackslash\hspace{0pt}}m{#1}}
\newcolumntype{R}[1]{>{\raggedleft\let\newline\\\arraybackslash\hspace{0pt}}m{#1}}

\title{Exploring the Role of Emotion Regulation Difficulties in the Assessment of Mental Disorders}

\author{{Rohan Kumar Gupta,  Rohit Sinha}\\
	Department of Electronics and Electrical Engineering,
Indian Institute of Technology Guwahati, Guwahati, India-781039 \\
	\texttt{(rohan\_kumar, rsinha)@iitg.ac.in} \\
}

\renewcommand{\shorttitle}{Emotion Regulation Difficulties Based Assessment of Mental Disorders}

\hypersetup{
pdftitle={A template for the arxiv style},
pdfsubject={q-bio.NC, q-bio.QM},
pdfauthor={David S.~Hippocampus, Elias D.~Striatum},
pdfkeywords={Conventional machine learning models, Representation learning},
}

\begin{document}
\maketitle

\begin{abstract}
Several studies have been reported in the literature for the automatic detection of mental disorders. It is reported that mental disorders are highly correlated. The exploration of this fact for the automatic detection of mental disorders is yet to explore. Emotion regulation difficulties (ERD) characterize several mental disorders. Motivated by that, we investigated the use of ERD for the detection of two opted mental disorders in this study. For this, we have collected audio-video data of human subjects while conversing with a computer agent based on a specific questionnaire. Subsequently, a subject's responses are collected to obtain the ground truths of the audio-video data of that subject. The results indicate that the ERD can be used as an intermediate representation of audio-video data for detecting mental disorders.
\end{abstract}


\section{Introduction}
\label{sec:introduction}

Traditionally, a self-reported questionnaire is used by clinicians for the initial screening of patients for mental disorders. To further enhance the screening process, the researchers have explored the use of additional modalities captured during the interaction of a clinician/computer agent with the subject under screening. These modalities include the recording of audio-video and/or physiological data of the session. The use of audio and video modalities is preferable for being non-intrusive as well as cost-effective.

Several automated systems are developed to assess several mental disorders, and a recent literature review of those can be found in~\citep{ReviewMHD}. In these reported works, the detection of mental disorders is addressed directly, i.e., without estimating any common attribute. On the other hand, there are some studies that focus on specific attributes that are common across those mental disorders and then exploit them in the assessment process. In the following, we review one widely used attribute referred to as emotion regulation difficulties (ERD).

Several emotions get induced in an individual while they attend to a situation. The intensity and polarity of induced emotions differ with the ability of an individual to regulate their emotions. As defined by Ross Thompson, emotion regulation refers to \emph{the extrinsic and intrinsic processes responsible for monitoring, evaluating, and modifying emotion reactions to accomplish one’s goals}~\citep{ERdefinition94}.
Many mental disorders are said to be characterized by ERD~\citep{ER_AS14Gross}. Several studies are reported for the measurement of ERD~\citep{r2,r3,r4}. These studies mainly focus on a self-report measure, namely, Difficulties in Emotion Regulation Scale (DERS)~\citep{DERSGratz04}. This measure is motivated for conceptualizing ERD as a  multidimensional construct. The DERS measures the ERD level of an individual through six dimensions that characterize:  (i) lack of emotional clarity (Clarity), (ii) non-acceptance of emotional responses (Non-acceptance), (iii) difficulty engaging in goal-directed behavior (Goals), (iv) difficulty in controlling impulsive behaviour (Impulse), (v) lack of emotional awareness (Awareness), and (vi) limited access to emotion regulation strategies (Strategies). Deficiencies in one or multiple dimensions of DERS indicate the level of difficulty an individual faces while regulating their emotions. In~\citep{DERSMDsymptoms09}, the authors have investigated the relationship between the subscales (i.e., dimensions) of DERS and several psychopathological symptoms. The severity levels of psychopathological symptoms is obtained through a self-report questionnaire, namely the Brief Symptom Inventory (BSI)~\citep{BSI1993}. It is reported that the DERS subscales and the severity levels of several psychopathological symptoms are highly correlated. The authors in~\citep{HallionDERS18} have computed the correlation coefficients among the DERS subscales and the severity levels of depression. It is noted that all DERS subscales and the severity levels of depression have a high correlation, except for the Awareness subscale. Similar correlation trends among the DERS subscales and the severity levels of PTSD is observed in~\citep{WeissCorrelations12}.

In recent times, the researchers have explored the use of psychophysiological signals towards further enhancing the diagnosis based on self-report measurement of ERD. In~\citep{DERSHRV17}, the authors have studied the association between heart rate variability (HRV) and DERS questionnaire. Their results indicate that DERS subscale scores and HRV are positively correlated. Other physiological factors considered in emotion regulation studies include facial corrugator electromyography responses~\citep{ERDfacial11}, respiration amplitude~\citep{ERDRA06}, and skin conductance level~\citep{ERDHRskincond15}. In~\citep{ERDfMRI02}, the authors have explored the association of functional magnetic resonance imaging (fMRI) data with ERD. The fMRI setup happens to be quite expensive, and the subjects are required to remain stationary for recording an unambiguous image~\citep{ERDfMRI16}. The collection of physiological and fMRI data is intrusive as well as costly. Instead, one could used audio and video data for estimating the ERD as these modalities are non-intrusive as well as more cost-effective.
To the best of our knowledge, no research has been reported on estimating the ERD using audio-video modality. Also, the role of ERD in assessing mental disorders is mainly investigated through self-report measures. The investigation about the same through audio-video data is yet to be explored.

Motivated by the above, in this work, we aim to estimate ERD using audio-video modality and explore its interaction with mental disorders. For that, we first present a methodology for the estimation of ERD\footnote[2]{We presented an initial work on estimating ERD using audio data in~\citep{ERD_TENCON21}.} on locally collected audio-video data recorded while subjects are conversing with an interactive system. Following that, the estimated ERD is used to assess two opted mental disorders, MDD and PTSD. The salient contributions of our present work are summarized below.

\begin{itemize}
     \item Creation of an audio-video database targeting the estimation of ERD, MDD, and PTSD.
    \item Proposed a methodology for automatic estimation of ERD using audio-video data. 
    \item Utilization of the estimated ERD for computing the severity levels of MDD and PTSD.
   
\end{itemize}

The rest of the paper is organized as follows.
The data collection is described in Section~II. Section~III presents the proposed methodology for the estimation of ERD. The experimental results are presented in Section~IV. Finally, the paper is summarised and discussed in Section~V.

\begin{figure}[t]
\begin{center}
\centerline{\includegraphics[width=8cm,height=8cm,keepaspectratio]{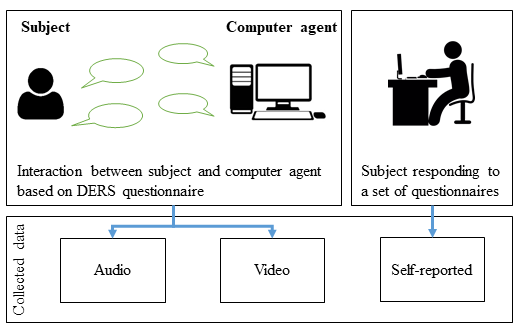}}
\end{center}
\vspace*{-4mm}
\caption{Schematic diagram outlining the protocol followed during the data collection.}
\label{datacollection}
\vspace*{8pt}
\end{figure}

\begin{figure}[t]
\begin{center}
\centerline{\includegraphics[width=9cm,height=9cm,keepaspectratio]{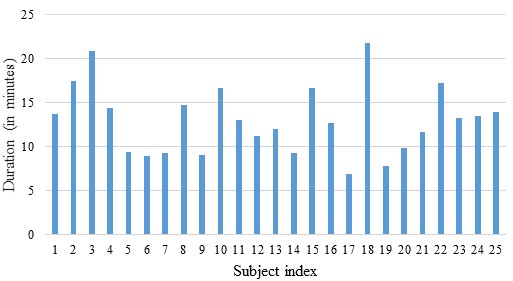}}
\end{center}
\vspace*{-6mm}
\caption{Subjectwise duration of the audio-video data in our collected dataset.}

\label{dataDis}
\vspace*{6pt}
\end{figure}

\section{The Data Collection}
 The schematic diagram of the data collection setup is shown in Figure~\ref{datacollection}.
 A detailed description of the data collection is provided in our recent paper~\citep{ERD_TENCON21}. For the sake of completeness, the brief details of the data collection are provided in the following. At first, the computer agent interacts with a subject through DERS questionnaire. The interaction is recorded in audio and video form. On completing the interaction with the computer agent, the subject fills out an online form comprising a set of questionnaires. The self-reported responses of a subject are used as the ground truth for their spoken and visual responses. In contrast to~\citep{ERD_TENCON21}, the present study also involves the self-reported responses to the Patient Health Questionnaire (PHQ-8)~\citep{PHQ8} and PTSD Checklist -- Civilian version (PCL-C)~\citep{PCLC1998} questionnaire that is used as the ground truths for the detection of MDD and PTSD, respectively. The threshold score of the PHQ-8 and PCL-C is 10 and 30, respectively. Since the publication of our previous work~\citep{ERD_TENCON21}, we have extended our dataset, which now comprises data from 25 subjects (18 males, 7 females), and their ages lie between 25 and 30 years. The approval for the data collection has been obtained from the Institutional Human Ethics Committee vide the reference IHEC/2022/RS/1 dated 21/06/2022. Figure~\ref{dataDis} shows the subject-wise duration of the spoken and visual responses to DERS questionnaire. It is noted that 14.82 seconds is the mean response duration to the questions. The collected dataset comprises 5.41 hours of spoken and visual responses corresponding to 25 subjects.

\begin{figure*}[t]
\begin{center}
\includegraphics[width=20cm,height=10cm,keepaspectratio]{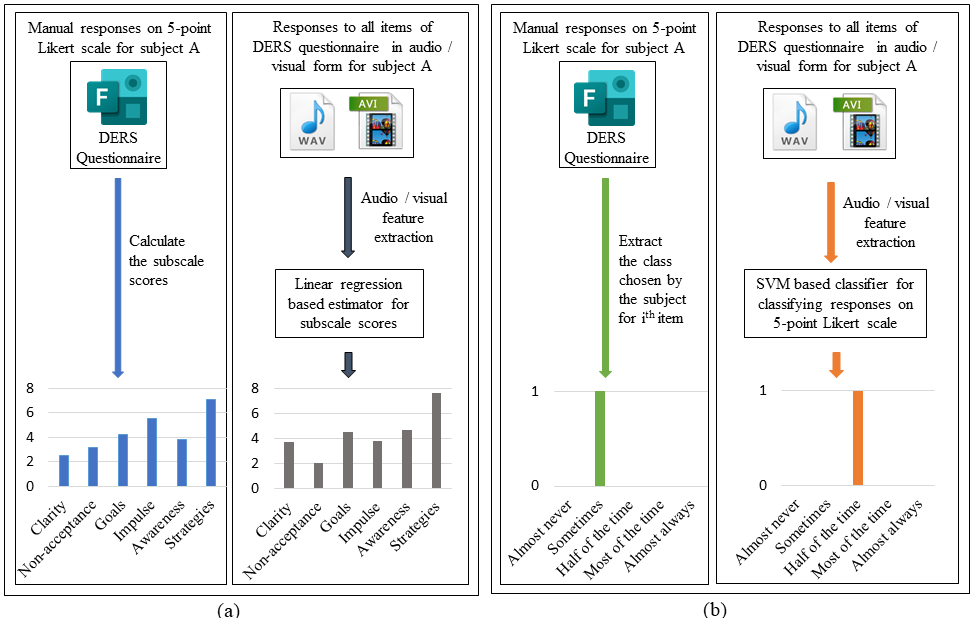}
\end{center}
\caption{Employed methodology for estimating (a) the DERS subscale scores, and (b) the DERS question response.}
\label{Methodology_MH_DI}
\end{figure*}

\section{Methodology}

The methodology employed for estimating the ERD using audio-video data of subjects' interaction in the context of DERS questionnaire is also inspired by a recent article~\citep{NakagawaQoL20} that discussed the quality of life estimation. We have performed two kinds of experiments for estimating the ERD, and those are described below:
\begin{itemize}
\item{{\bf DERS subscale score estimation}: In this case, the DERS subscale scores of a subject are estimated directly from the subject’s audio-video responses to the items in the corresponding subscales (\emph{regression} problem).}
\item{{\bf DERS questionnaire-response estimation}: It involves the estimation of a subject’s response to each item in the DERS questionnaire from their audio-video data corresponding to that item (\emph{classification} problem).}
\end{itemize}

The overall methodology of the above-mentioned experiments is depicted in Figure~\ref{Methodology_MH_DI}.
For video modality, a subset of the feature set provided in~\citep{AVEC16} for the depression sub-challenge is utilized, which includes action units (AUs) and gaze direction estimate for both eyes (eye-gaze). These video features are extracted using the OPENFACE toolbox~\citep{OPENFACE}. The feature set used for the audio modality is similar to discussed in our recent paper~\citep{ERD_TENCON21}. The fusion of the two audio and video modalities is performed by concatenating the features of the individual modalities (feature-level fusion).
This study uses a random forest regressor from the sklearn library for the regression task. The number of trees, a hyper-parameter, in the random forest is set to 10. For the classification task, an SVM having an RBF kernel is used from the sklearn library.
The rest of the experimental setup is similar to and discussed in our recent paper~\citep{ERD_TENCON21}.

\section{Results}
In this section, we present the performance evaluation of the ERD estimation and its efficacy in predicting the two opted mental disorders (MDD and PTSD).

\subsection{ERD estimation performance}
In the following, we present the assessment of the ERD estimation performance. Table~\ref{tableDERS_EE} shows the RMSE values for DERS subscale score estimation in the case of direct and indirect estimation approaches for three modalities. The direct approach refers to the  DERS subscale score estimation experiment, whereas the subscale scores obtained through DERS questionnaire-response estimation experiment refer to the indirect approach. On comparing the two approaches across the modalities, it is noted that all subscales in the direct approach yield better performance for all three modalities, except the Awareness subscale. The indirect approach provides better performance only for the Awareness subscale in all modalities, except video. Thus, the direct approach outperforms the indirect one in an overall sense.

\begin{table}[b]
\begin{center}
\caption{RMSE values for DERS subscale score estimation using direct and indirect approaches corresponding to audio (A), video (V), and audio-video (A-V) modalities.  }
\begin{tabular}{c c c c c c c}
\hline
\hline
\multicolumn{1}{l}{\textbf{Subscale}} & \multicolumn{3}{c}{\textbf{Indirect approach}} & \multicolumn{3}{c}{\textbf{Direct approach}}\\
\cmidrule(lr){2-4}
\cmidrule(lr){5-7}
\multicolumn{1}{r} {\textbf{Modality}$\rightarrow$}& \textbf{A} & \textbf{V} & \textbf{A-V} & \textbf{A} & \textbf{V} & \textbf{A-V}\\
\hline

\multicolumn{1}{l}{Clarity} & 2.31  & 2.78 & 2.47 & 2.27 & 2.74 & 2.33 \\
\multicolumn{1}{l}{Non-acceptance} & 5.35 & 5.57 & 5.37 & 5.33 & 5.40 & 4.96\\
\multicolumn{1}{l}{Goals} &	5.10 & 5.04 & 5.28 & 5.09 & 4.91 & 4.77\\
\multicolumn{1}{l}{Impulse} & 4.94 & 4.83 & 5.22 & 4.25 & 4.45 & 4.18\\
\multicolumn{1}{l}{Awareness} &	4.27 & 4.78 & 4.25 & 4.70 & 4.72 & 5.22\\
\multicolumn{1}{l}{Strategies} & 6.14 & 6.38 & 6.30 & 5.33 & 5.30 & 5.19\\

\hline
\hline
\end{tabular}
\label{tableDERS_EE}
\end{center}
\end{table}

\begin{table}[t]
\begin{center}
\caption{Performance of estimating MDD and PTSD severity levels using self-reported data}
\begin{tabular}{c c c}

\hline
\hline
			\multicolumn{1}{c}{\textbf{Mental Disorder}}& \multicolumn{1}{c}{\textbf{MAE}} & \multicolumn{1}{c}{\textbf{RMSE}}\\
\hline
MDD & 1.98 & 2.57\\
PTSD & 7.66 & 9.28\\

\hline
\hline
\end{tabular}
\label{tableRoleERDselfreport}
\end{center}
\vspace*{-2mm}
\end{table}

\begin{table}[t]
	\caption{\fontsize{8}{9}\selectfont Comparing the performance of MDD and PTSD severity levels estimation via ERD assessed using audio-video data as well as bypassing it. }
	\begin{center}
		\begin{tabular}{| c | c | c | c | c |}
			\hline			
			\multicolumn{5}{|c|}{\textbf{MDD}} \\
			\hline
			\hline
		\cline{1-5}
		\multicolumn{1}{|c}{\textbf{Modality (Fusion)}}& \multicolumn{2}{|c}{\textbf{Bypassing ERD}} &\multicolumn{2}{|c|}{\textbf{Via ERD}}\\
			\cline{2-5}
			\multicolumn{1}{|c}{}& \multicolumn{1}{|c}{\textbf{MAE}} & \multicolumn{1}{|c}{\textbf{RMSE}} &\multicolumn{1}{|c}{\textbf{MAE}}&\multicolumn{1}{|c|}{\textbf{RMSE}}\\
            \hline
			\hline
			Audio & \multicolumn{1}{c}{3.73} &	\multicolumn{1}{|c|}{4.65}& \multicolumn{1}{c}{3.98} &	\multicolumn{1}{|c|}{4.86}\\
			Video & \multicolumn{1}{c}{4.42} &	\multicolumn{1}{|c|}{5.26} & \multicolumn{1}{c}{4.03} &	\multicolumn{1}{|c|}{5.08}  \\
			Audio-Video (Feature-level) & \multicolumn{1}{c}{4.02} &	\multicolumn{1}{|c|}{4.96} & \multicolumn{1}{c}{4.22} &	\multicolumn{1}{|c|}{5.08} \\
			\hline
		     \hline			
		     \multicolumn{5}{|c|}{\textbf{PTSD}} \\
			\hline
			\hline
		\cline{1-5} 
         \multicolumn{1}{|c}{\textbf{Modality (Fusion)}}& \multicolumn{2}{|c}{\textbf{Bypassing ERD}} &\multicolumn{2}{|c|}{\textbf{With ERD}}\\

			\cline{2-5}
			\multicolumn{1}{|c}{}& \multicolumn{1}{|c}{\textbf{MAE}} & \multicolumn{1}{|c}{\textbf{RMSE}} &\multicolumn{1}{|c}{\textbf{MAE}}&\multicolumn{1}{|c|}{\textbf{RMSE}}\\
         \hline
			\hline
			Audio & \multicolumn{1}{c}{8.67} &	\multicolumn{1}{|c|}{10.47} & \multicolumn{1}{c}{10.42} &	\multicolumn{1}{|c|}{12.50}\\
			Video & \multicolumn{1}{c}{10.88} &	\multicolumn{1}{|c|}{12.96}& \multicolumn{1}{c}{10.95} &	\multicolumn{1}{|c|}{13.04} \\
			Audio-Video (Feature-level) & \multicolumn{1}{c}{8.95} &	\multicolumn{1}{|c|}{10.92}& \multicolumn{1}{c}{11.46} &	\multicolumn{1}{|c|}{13.49}\\
			\hline
			
		\end{tabular}
		\label{tableRoleERD}
	\end{center}
\end{table}

\subsection{Role of ERD in assessing MDD and PTSD }
In the following, we present the results of the experiments performed to establish the role of ERD for the assessment of  MDD and PTSD severity levels. In these experiments, a training instance corresponds to a subject's six DERS subscale scores derived by considering that subject specific self-reported data. The ground truth is the severity level of MDD/PTSD which is obtained through the self-reported responses of that subject. 
A regression model is used to learn the relationship between six DERS subscale scores and the severity levels of MDD/PTSD.
The trained regression model is evaluated, using the leave-one-subject-out method, with test inputs derived from: (i) DERS subscale scores obtained through self-reported responses, and (ii) DERS subscale scores estimated through the direct approach of the ERD estimation. Table~\ref{tableRoleERDselfreport} shows the estimation errors of MDD and PTSD severity levels for the test instances obtained through the self-reported responses. For validating the case (ii), another experiment is performed in which the severity levels of MDD/PTSD are estimated directly (i.e., without estimating ERD) from audio and video responses using leave-one-subject-out method. For the same experiment, AVEC16-audio is used as the audio feature set. Whereas, the video feature set contains AUs and eye-gaze. A random forest regressor from the sklearn library is used for the regression task. The number of trees, a hyper-parameter, in the random forest is set to 10.

Table~\ref{tableRoleERD} shows the errors for estimating the MDD and PTSD severity levels via ERD assessed using audio-video data as well as bypassing it. It can be observed from the table that the performances of MDD and PTSD severity levels estimation via assessed ERD have turned out closely similar to those of bypassing it. Therefore, we can state that the ERD provides a viable means for the assessment of opted mental disorders. On comparing the MDD and PTSD severity level estimation performances reported in Table~\ref{tableRoleERD} with Table~\ref{tableRoleERDselfreport}, the noted degradation are attributed to the errors in estimating DERS subscale scores in the former case. Thus, there is a scope of further enhancing the severity levels estimation of MDD and PTSD with improved ERD estimation.

\section{ Summary and Discussion}
This study is aimed at estimating the ERD using audio-video modality and exploring its interaction with mental disorders. For that, we first presented a methodology for the estimation of ERD on locally collected audio-video data recorded while the subjects were conversing with an interactive system. Following that, the estimated ERD is used for the assessment of two opted mental disorders, MDD and PTSD.

For the estimation DERS subscale scores, two approaches, referred to as direct and indirect, are investigated. Comparing those two approaches across three modalities, we observed that the direct approach provides better estimation performance for all DERS subscales, except for the Awareness subscale. Thus, the direct approach outperforms the indirect one in an overall sense.

The estimated ERD with the direct approach is further utilized for the assessment of MDD and PTSD severity levels. The performance of MDD/PTSD severity level estimation involving ERD is compared with another experiment that estimates the severity level of MDD/PTSD directly from audio-video responses without estimating ERD. The performance in the two cases are found to be quite similar. Therefore, we conclude that the ERD can be used as an intermediate representation of audio-video data to estimate the severity levels of MDD/PTSD.

The salient limitations in this study are discussed as follows.
In this study, the self-reported data is used as the ground truth, and that may have reporter's bias. Towards addressing that, the help of an expert, says a psychologist, can be sought for obtaining the ground truth for the audio-video data.
Further, we did not make use of deep learning models as those may not get trained reliably on a small-sized dataset created for this study. So in the future, it will be worthwhile to replicate the key findings of this study on a larger dataset.


\bibliographystyle{unsrtnat}
\bibliography{references} 

\end{document}